\documentclass[twocolumn,showpacs,pra]{revtex4}

\usepackage[colorlinks=true,citecolor=blue,linkcolor=magenta]{hyperref}

\usepackage{amsmath}
\usepackage{amssymb}
\usepackage{txfonts}
\usepackage{graphicx}
\usepackage{epstopdf}
\usepackage{dcolumn}
\usepackage{bm}
\begin{document}
\title{Detection of light-matter interaction in the weak coupling regime by quantum light}
\author{Qian Bin}
\author{Xin-You L\"{u}}\email{xinyoulu@hust.edu.cn}
\author{Li-Li Zheng}
\author{Shang-Wu Bin}
\author{Ying Wu}\email{yingwu2@126.com}
\affiliation{School of Physics, Huazhong University of Science and Technology, Wuhan, 430074, P. R. China}
\date{\today}

\begin{abstract}
``Mollow spectroscopy'' is a photon statistics spectroscopy, obtained by scanning the quantum light scattered from a source system. Here, we apply this technique to detect the weak light-matter interaction between the cavity and atom (or a mechanical oscillator) when the strong system dissipation is included. We find that the weak interaction can be measured with high accuracy when exciting the target cavity by quantum light scattered from the source in the halfway between the central peak and each side peak. This originally comes from the strong correlation of the injected quantum photons. In principle, our proposal can be applied into the normal cavity quantum electrodynamics (QED) system described by JC model and optomechanical system. Furthermore, it is state-of-the-art for experiment even when the interaction strength is downed to a very small value.
\end{abstract}

\maketitle
\section{Introduction}
One of the central topics of modern optics is that the investigations of light-matter interaction~\cite{
ref1}. Studies on light-matter interaction in a variety of systems have been extended to the strong coupling regime in recent years~\cite{ref2,ref2-1,ref3,ref4}. These investigations are very useful for the implementation of coherent manipulations in quantum information science, and have also potential applications in the development of practical quantum devices. Even so, it is still difficult for realizing the strong lighter-matter interaction in some systems, such as the cavity optomechanical system (OMS). Cavity optomechanics is an emerging field, which explores the interaction between electromagnetic radiation and mechanical resonator motion, and has progressed enormously in recent years~\cite{ref11}. These achievements including the realization of squeezed light~\cite{ref12,ref13,ref14}, precision measurements~\cite{ref15,ref16}, demonstration of optomechanically induced transparency and fundamental tests of quantum mechanics~\cite{ref17,ref18}. Recently, it has also been present that the single-photon strong coupling can be realized in an OMS even it is originally in the weak coupling regime~\cite{ref19,ref21}. However, the strong interaction between well-coupled optical mode and mechanical oscillator in the OMS is not easy to achieve, the light-matter interaction under weak coupling regime is still a field worth studying~\cite{ref20,ref22,ref23,ref24,ref24-1,ref24-2}.

Recently, it has been proposed that, in a quantum system that consists of two linearly coupled harmonic oscillators and weakly interacting excitations, the weak Kerr nonlinearities can be detected with high precision even when the system is in the strongly dissipative environments. The main reason for the realization of this detection is that the use of a new spectroscopic technique-``Mollow spectroscopy''~\cite{ref30}. It is a theoretical concept of the photon statistics spectroscopy~\cite{ref30-1}, obtained by scanning the output of resonance fluorescence from the source into a target system. This method of detecting weak Kerr nonlinearities in the quantum system is different from the weak value measurements. Note that the weak value measurement could be used to amplify some weak signals and observe nonclassical phenomena. It has been studied in different systems~\cite{ref25,ref26,ref27,ref28,ref29}. Moreover, it is well-known that the energy level splitting can generate the ladder of dressed states in a cavity QED system described by JC model. But the ladder is disappeared when the system is in the weak coupling regime. The reason is that high system dissipation leads to the separation of splitting are covered by the widths of dressed states. Therefore the detection of weak interaction in this system becomes a difficult job. Similarly, for an OMS that consists of a cavity mode coupled to a mechanical resonator, there is a small shift of the emission peak when the system under the strong coupling regime. It has also been discovered that the coupling strength can be observed by measuring the shift of the peak\cite{ref19,ref31,ref41,ref42}. However, the shift will be covered when the system is in the weak coupling regime. So this method is unsuitable for measuring weak interaction in OMS. Then one question arises naturally. Whether this new photon statistics spectroscopy can be used to detect the weak interaction between light and matter in cavity QED system and OMS.

Motivated by the above question. In this paper we study the responses of the cavity QED system and OMS in the weak coupling regime to the input quantum field from source system. Here the source is made of a two-level atom driven by the classical light fields. The output field of source is called as ``quantum light'', which could be scanned onto the target cavity to form new emission spectrum and statistics spectroscopy-``Mollow spectroscopy'' \cite{ref30,ref31-1,ref32,ref33}. Here both the population and the statistical property of source system are transferred to the target cavity, but with some deviations due to the presence of interaction in the target system. Moreover, the photon statistics spectroscopy has higher sensitivity than the emission spectrum. We thus apply the deviation of photon statistics spectroscopy to probe the weak interaction in cavity QED system and OMS when the strong system dissipation is also included. We find that, in the weak coupling regime, the interaction strength can be observed with high precision. Particularly, the weak detection is still state-of-the-art for experiment even when the interaction strength is a very small value.

Our paper is organized as follows. In Sec.\,II, we discuss the cavity QED system driven by the classical and quantum light fields, and present the detection of the interaction between cavity and atom in the weak coupling regime by quantum light. In Sec.\,III, we introduce the OMS that consists of a single-mode cavity weakly coupled to a mechanical resonator, and present the detection of the interaction between cavity and mechanical modes by the same quantum light. In Sec.\,IV, we give discussions for the experimental realization in our proposal. In Sec.\,V, we give conclusions of our work.

\begin{figure}
  \centering
  \includegraphics[width=8cm]{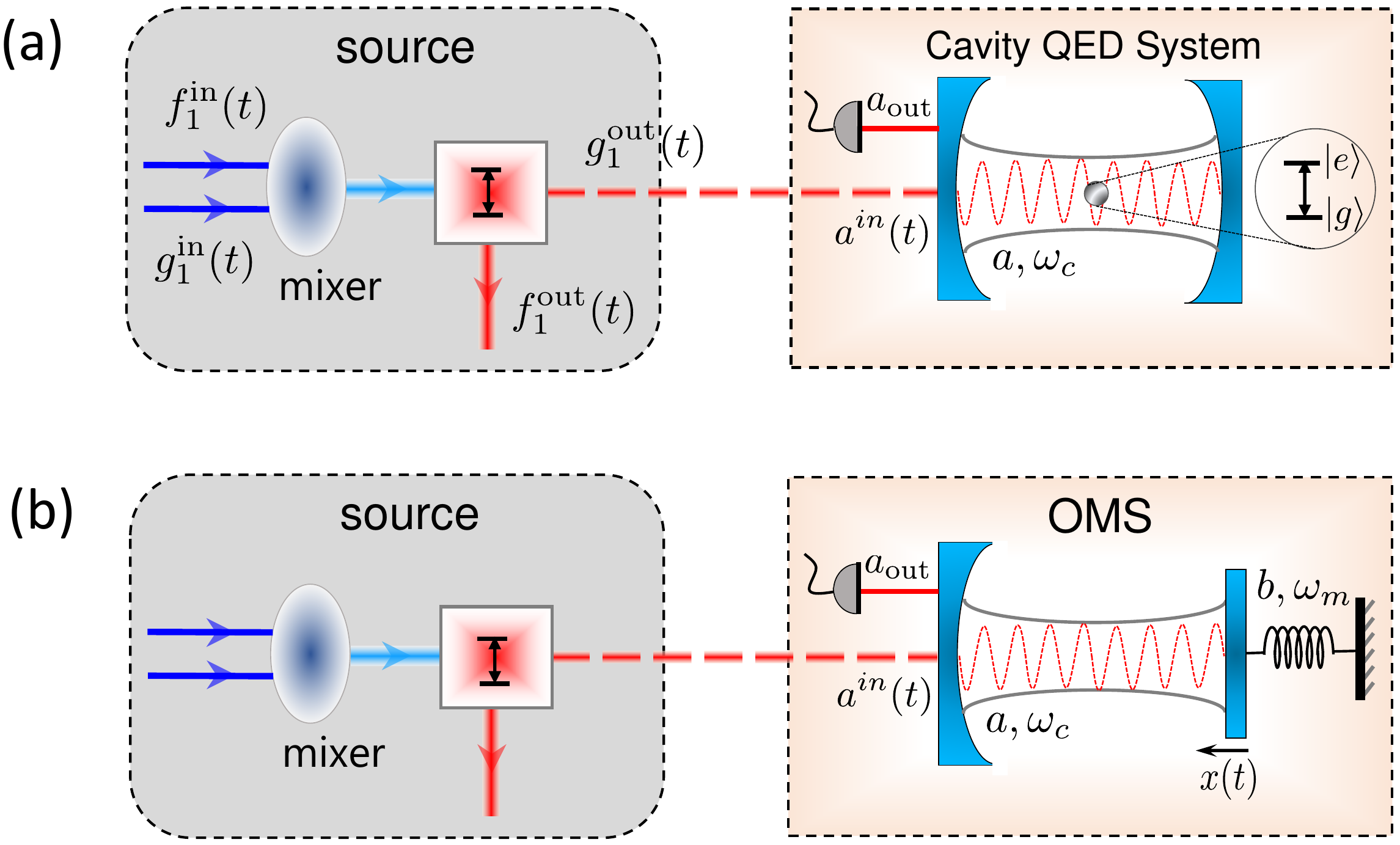}\\
  \caption{ Schematics of the studied systems. (a) The cavity QED system consisting of a two-level atom coupled to a single-mode cavity driven by the emission of a quantum source; (b) The optomechanical system is excited by the emission of the same quantum source. Here, the quantum source is made of a two-level atom driven by classical light fields, $f_1^{in}(t)$ and $g_1^{in}(t)$ are two input channels for exciting the source two-level atom, $f_1^{out}(t)$ and $g_1^{out}(t)$ are the output channels of the source system, $a^{in}(t)$ is the input channel of the target.}\label{1}
\end{figure}
\section{Detection of weak interaction in JC model by quantum light}
\begin{figure}
  \centering
  \includegraphics[width=7cm]{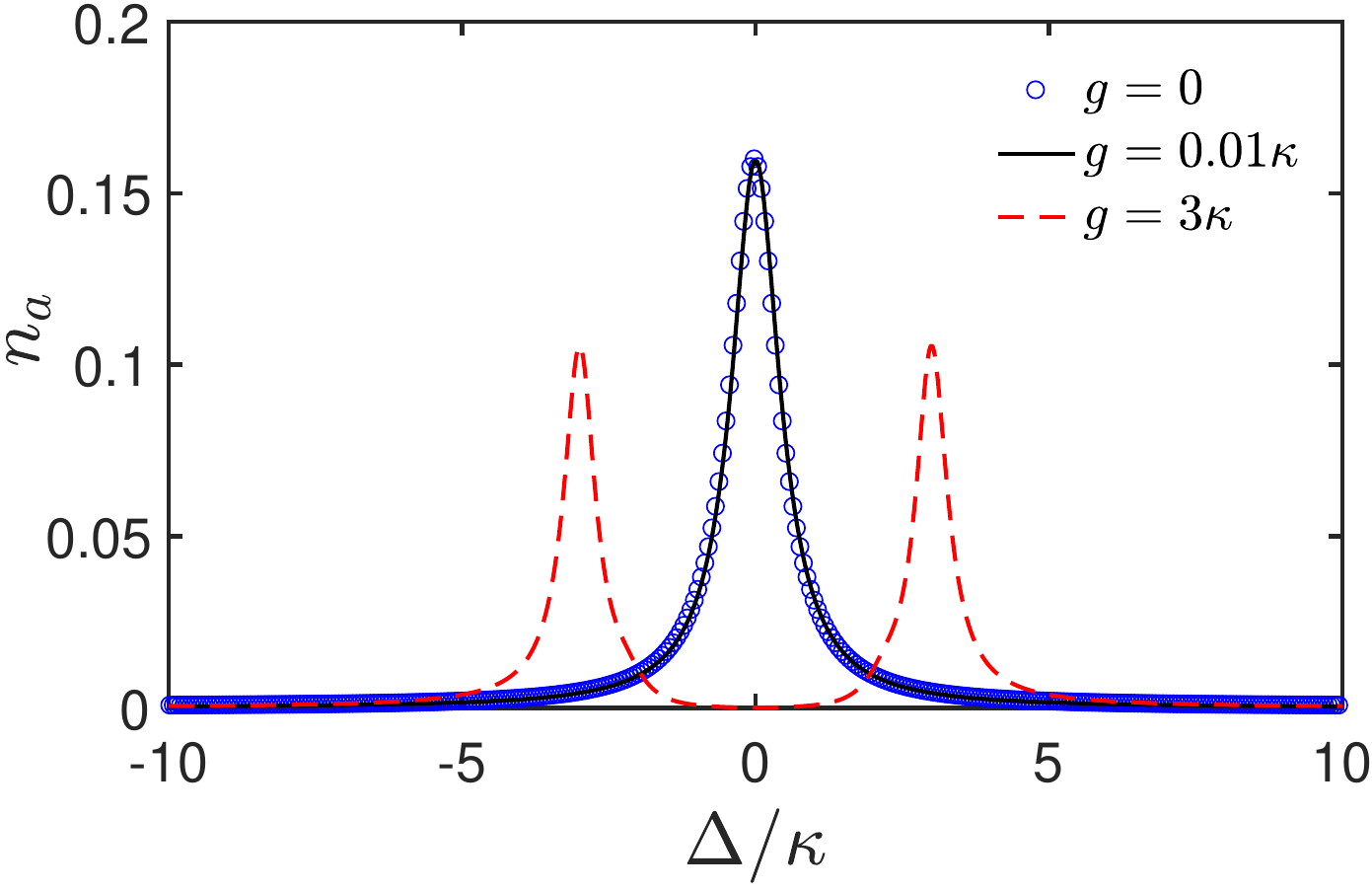}\\
  \caption{ Cavity mean photon number $n_a$ versus cavity-field detuning $\Delta$ when JC model is driven by classical field. The blue circles, black solid line and red dashed line in panels correspond to the atom-cavity coupling strength $g=0$, $g=0.01\kappa$ and $g=3\kappa$, respectively. The system parameters used here are: $\gamma=0.001\kappa$ and $\Omega=0.6\kappa$.}\label{2}
\end{figure}

In cavity QED system, the clear energy splitting arose from the generation of dressed states can be obtained when a cavity strongly coupled to an atom, as displayed in Fig.~\ref{2}. The separation of splitting could be used to detect the interaction strength between cavity and atom in the system. However, when the system is in the weak coupling regime, the obvious splitting will disappear. This is because the weak interaction between atom and cavity leads to the separation of splitting is covered by the widths of dressed states. So the detection of weak coupling in this system by the classical light is difficult.

\begin{figure*}
  \centering
  \includegraphics[width=16cm]{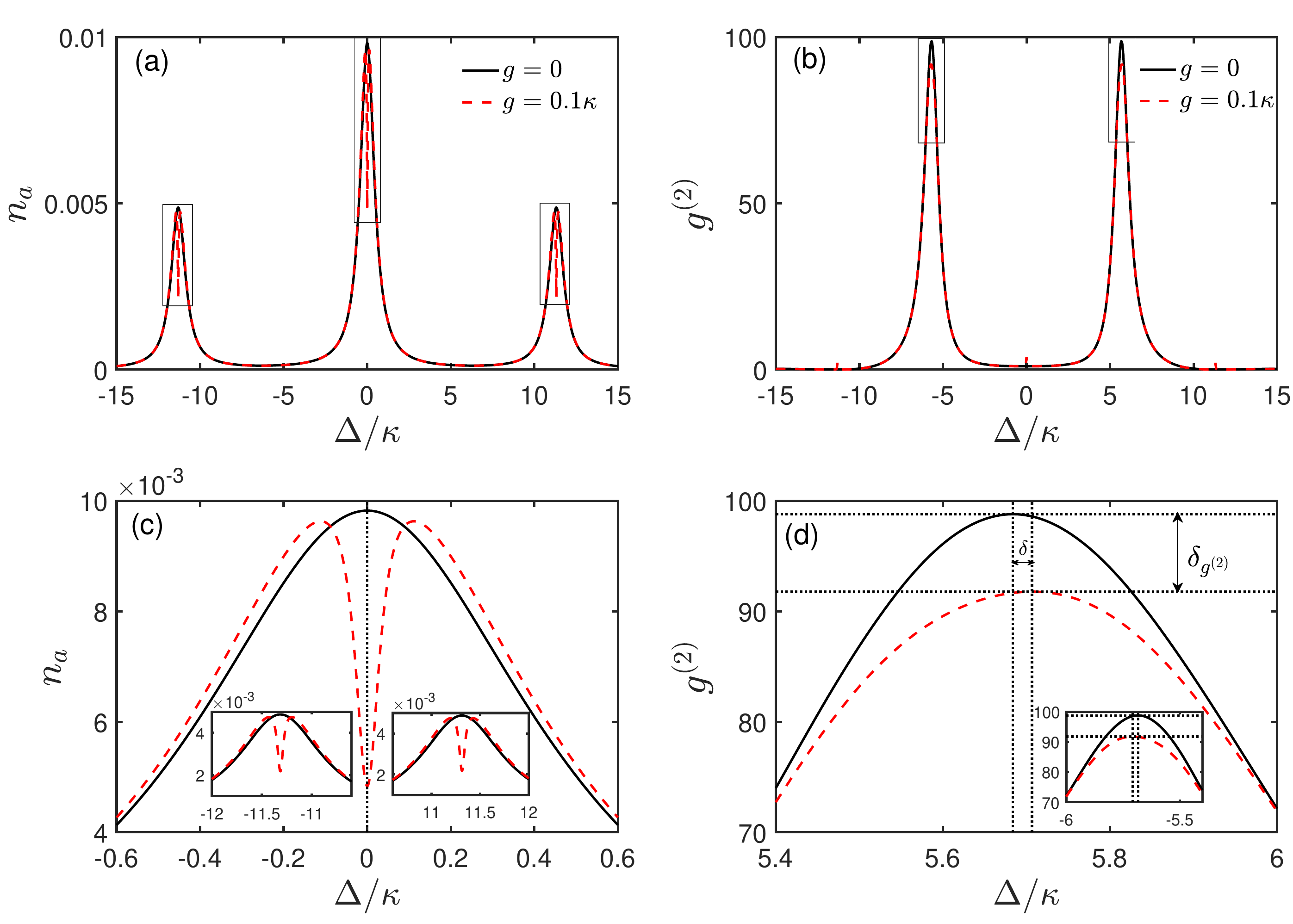}\\
  \caption{ Plot the photon emission spectrum (a) and photon statistics spectrum (b) when the Mollow triplet is scanned into the JC model. The black solid line and red dashed line in panels correspond to the atom-cavity coupling strength $g=0$ and $g=0.1\kappa$, respectively. (c) Enlarged view of the spectral region of middle peak delimited by square in (a). Insets: Enlarged view the regions of the left and right emission peaks in (a). These spectrum lines show the splittings in the three emission peaks. (d) Enlarged view of the spectral regions of peaks delimited by squares in (b), here the left peak is enlarged in the inset. $\delta$ represents the shift of statistics peak when $g=0.1\kappa$, and $\delta_{g^{(2)}}$ is the difference of the peak with $g=0.1\kappa$ and $g=0$. The system parameters used here are: $\gamma_s=0.02\kappa$, $\gamma=0.001\kappa$, $\Omega=8\kappa$, $\mu_1=0.5$ and $\mu_2=0.5$.}\label{3}
\end{figure*}
\begin{figure}
  \centering
  \includegraphics[width=7cm]{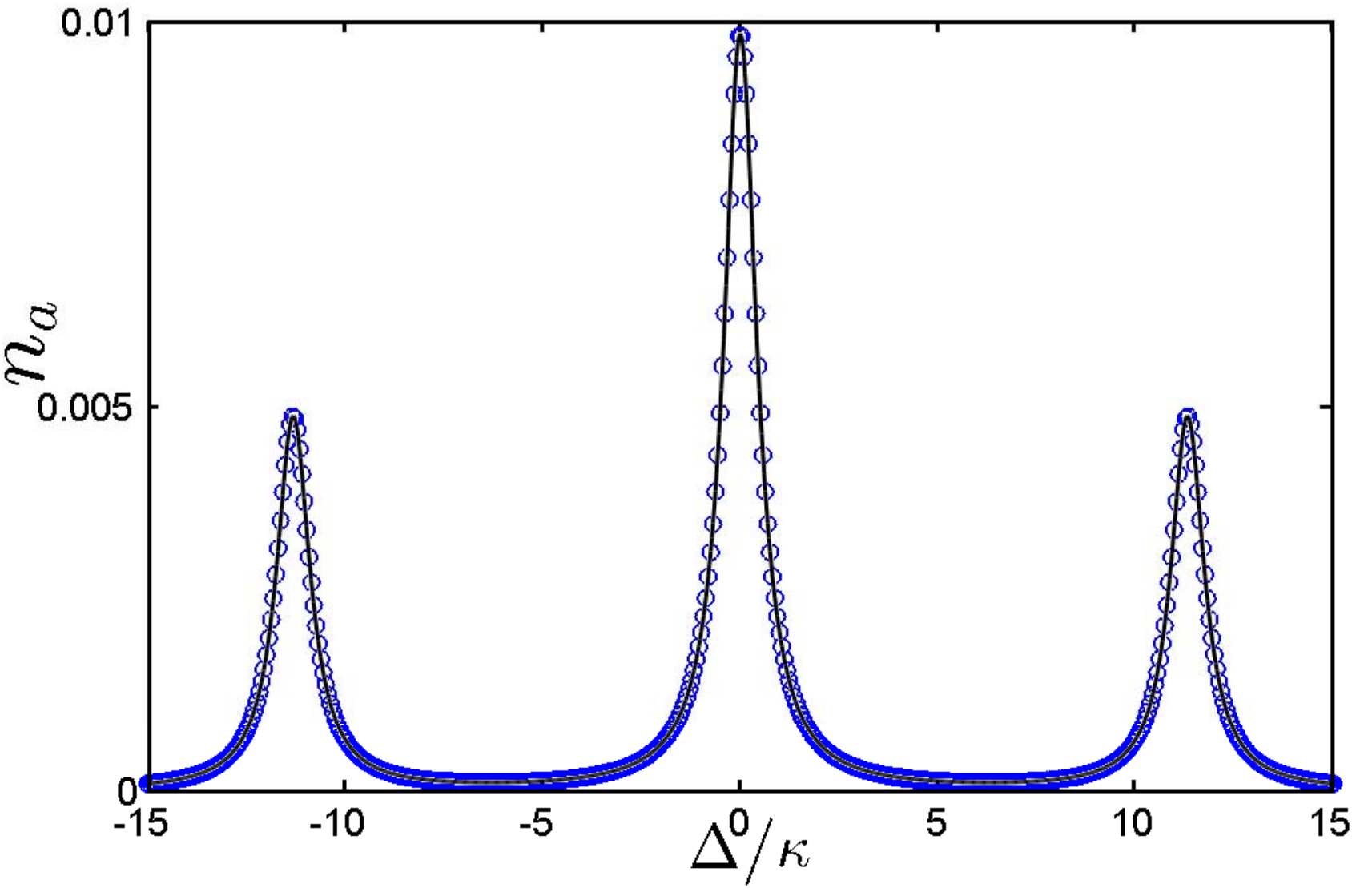}\\
  \caption{ Plot the photon emission spectrum $n_a$ of steady-state for $g=0$ obtained from Eq.(10) (blue circles) and exact numerical calculation (black solid curve). The other system parameters used here are the same as in Fig.~\ref{3}.}\label{4}
\end{figure}
As shown in Fig.~\ref{1}(a), we now consider a cavity QED system that consists of a two-level atom weakly coupled to a single-mode cavity driven by quantum source. The source system is made of a two-level atom driven by classical light fields. We assume that there are two input channels $f_1^{in}(t)$ and $g_1^{in}(t)$ for the source (with weights $\mu_1$ and $\mu_2$,  $\mu_1+\mu_2=1$) and only one input channel $a^{in}(t)$ for the target. $g_1^{in}(t)$ represents the vacuum field. Thus the source Hamiltonian is given by~\cite{ref34,ref35,ref36,ref37,ref39,ref40}
\begin{equation}\label{eq1}
H_s=\omega_s\ \sigma_s^\dag \sigma_s+\sqrt{\mu_1}\Omega \sigma_s e^{i\omega_L t}+\sqrt{\mu_1}\Omega^\ast \sigma_s^\dag e^{-i\omega_L t}.
\end{equation}
For the target system, which is the prototypical Jaynes-Cummings Hamiltonian
\begin{equation}\label{eq2}
H_{t_1}=\omega_c a^\dag a+\omega \sigma^\dag \sigma+g(\sigma^\dag a+\sigma a^\dag).
\end{equation}
Here, $\omega_L$ and $\Omega$ are frequency and intensity of the laser field, respectively. $a$ $(a^\dag)$ represents the annihilation (creation) operator of the cavity mode with resonant frequency $\omega_c$. $\sigma_s$ denotes the lowering operator of the source two-level atom with transition frequency $\omega_s$. $\sigma$ denotes the lowering operator of the target two-level atom with transition frequency $\omega$, and $g$ is atom-cavity coupling strength. The target system is excited by the output light field from source system, the main requisite is that the dynamics of source system is not affected by the presence of the target. Therefore, we consider coupling the source and target systems via a thermal bath. The dynamics of the coupled system is tackled in the framework of cascaded quantum system. Here, the output field of source is set as the input field of target via equations of motion, and there is no back action from the target. The coupling regime in cascaded quantum system involving the dissipative mediated excitation process, such a coupling is made with the decay of system. Under these conditions, we transform the system into a frame rotating with $ \omega_L$ to remove the time dependence. We thus derive the master equation
\begin{align}\label{eq3}
\frac {d\rho}{dt}=&i[\rho,H_s'+H_{t_1}']+\gamma_s\mathcal{L}[\sigma_s]+\kappa\mathcal{L}[a]
+\gamma\mathcal{L}[\sigma]\nonumber\\
&-\sqrt{\mu_2 \gamma_s \kappa} \{[a^\dag,\sigma_s \rho]+[\rho \sigma_s^\dag,a]\},
\end{align}
where
\begin{align}\label{eq4}
H_s'=\Delta_s\sigma_s^\dag \sigma_s+\sqrt{\mu_1}\Omega\sigma_s+\sqrt{\mu_1}\Omega^\ast\sigma_s^\dag,\\
H_{t_1}'=\Delta a^\dag a+\Delta_a \sigma^\dag\sigma+g(\sigma^\dag a+\sigma a^\dag),
\end{align}
and the superoperator $\mathcal{L}$ express as
\begin{equation}\label{eq6}
\mathcal{L}[\emph{O}]=\frac {1}{2}(2\emph{O}\rho\emph{O}^\dag-\rho\emph{O}^\dag\emph{O}-\emph{O}^\dag\emph{O}\rho).
\end{equation}
Here, $\gamma_s$ is the emission rate of the source two-level atom, $\gamma$ and $\kappa$ are decay rates of the two-level atom and cavity in the target system, respectively. $\Delta_s=\omega_s-\omega_L$, $\Delta=\omega_c-\omega_L$ and $\Delta_a=\omega-\omega_L$ are detunings with respect to the external driving field. $\sqrt{\mu_2 \gamma_s \kappa}$ represents the dissipative coupling strength between source and target systems.

\begin{figure*}
  \centering
  \includegraphics[width=16cm]{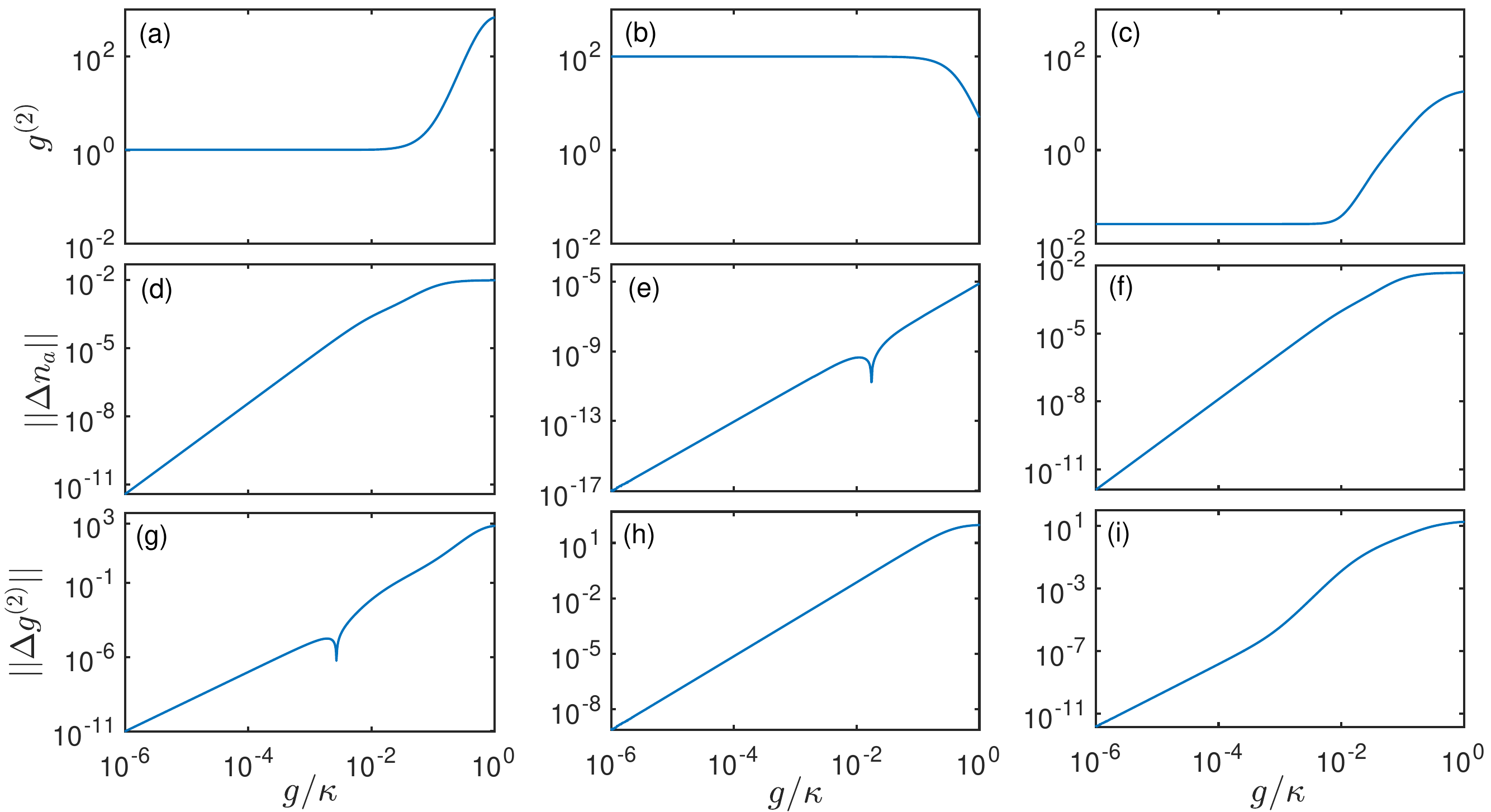}\\
  \caption{ The equal-time second-order photon correlation function $g^{(2)}$ (the first row), the norms $||\Delta n_a||$ (the second row) and  $||\Delta g^{(2)}||$ (the third row) of the target cavity versus $g/\kappa$ for different input types of quantum lights. The first, second and third columns correspond to quantum lights from the central peak, the emission halfway between the central peak and right peak, and the right peak, respectively. The system parameters used here are: $\gamma_s=0.02\kappa$, $\gamma=0.001\kappa$, $\Omega=8\kappa$, $\mu_1=0.5$ and $\mu_2=0.5$.}\label{5}
\end{figure*}

We consider driving the source two-level atom in the Mollow regime of a spectral triplet. There are various photon correlation types when choosing driving fields with different frequencies~\cite{ref39,ref40}. For instance, there are three peaks in the emission spectrum of the source system, which show three forms of photon correlations, i.e., antibunched, bunched and superbunched. Photons from the central peak are bunched, those from the side peaks are antibunched, and those from the emission halfway between the central peak and each side peak are superbunched. Quantum light from Mollow triplet can be scanned over the target cavity to form new emission spectrum and photon statistics spectroscopy-``Mollow spectroscopy", as displayed in Figs.~\ref{3}. The black solid and red dashed lines versus $g=0$ and $g=0.1\kappa$, respectively. Here, both the population and the statistics of the Mollow regime are transferred to the target cavity. Although these spectrums are still symmetrical about $\Delta=0$,  the mean photon number and equal-time second-order photon correlation function can occur some deviations. The reason for this behavior is the presence of the coupling between the cavity and atom in JC model. Enlarging these regions of spectrum peaks in Fig.~\ref{3}(a), we see three clear splittings in Fig.~\ref{3}(c). However, these splittings in practice could be covered by the width of the spectrum line due to weak interaction and high dissipation. From Fig.~\ref{3}(b), it is seen that the black and red curves are basically coincident. Actually, there is a shift $\delta$ between the peaks on an enlarged view as in Fig.~\ref{3}(d), and the left and right peaks have same deviations. Comparing the Figs.~\ref{3}(c) and ~\ref{3}(d), it can be seen that the difference of photon statistics is much larger than that of photon population in the same parameter regime, caused by the quantum character of system. Here, the deviation of photon statistics is not covered by the width of the spectrum line even when the system is in the weak coupling regime. Note that the deviation of photon population is not marked in Fig.~\ref{3}(c) because its value is too small.

To understand the dynamics of the system more clearly, we obtain the exact solution of photon population under the case with $g=0$ (the exact solution of equal-time second-order photon correlation function $g^{(2)}$ can also be obtained but its expression is more bulky)~\cite{ref32},
\begin{equation}\label{eq7}
n_a=\frac{16\Omega^2\gamma_s\mu_1\mu_2 A}{B(4\Delta^2 + \kappa^2)(8\mu_1\Omega^2 + \gamma_s^2)(4\Delta^2 + \gamma_s^2 + 2\gamma_s\kappa + \kappa^2)},
\end{equation}
where
\begin{align}\label{eq8}
B=&16\Delta^4 +256\Omega^4\mu_1^2+ 4\Delta^2(5\gamma_s^2 + 6\gamma_s\kappa + 2\kappa^2)\nonumber\\
&+32\Omega^2\mu_1(2\gamma_s^2 + 3\gamma_s\kappa + \kappa^2- 4\Delta^2) \nonumber\\ &+(\gamma_s+\kappa)^2(4\gamma_s^2+4\gamma_s\kappa+\kappa^2),
\end{align}
and $A=A_{1}+A_{2}+A_{3}$,
\begin{align}\label{eq9}
A_1=&64\kappa\Delta^6 + 16\Delta^4(8\mu_1\Omega^2(2\gamma_s - \kappa) + 6\gamma_s^2\kappa + 8\gamma_s\kappa^2 + 3\kappa^3),\\
A_2=&32\mu_1\Delta^2\Omega^2(16\Omega^2\mu_1(\gamma_s + \kappa) + 8\gamma_s^3 + 23\gamma_s^2\kappa + 16\gamma_s\kappa^2 + 2\kappa^3)\nonumber \\
&+ 4\kappa\Delta^2(9\gamma_s^4 + 28\gamma_s^3\kappa + 32\gamma_s^2\kappa^2 + 16\gamma_s\kappa^3 + 3\kappa^4),\\
A_3=&8\mu_1\kappa\Omega^2 (4\gamma_s^4 + 16\gamma_s^3\kappa + 23\gamma_s^2\kappa^2 + 14\gamma_s\kappa^3 + 3\kappa^4)\nonumber\\
&+\kappa(\gamma_s+\kappa)^2(2\gamma_s+\kappa)(2\gamma_s^3+5\gamma_s^2\kappa+4\gamma_s\kappa^2+\kappa^3)\nonumber\\
&+128\Omega^4\kappa^2\mu_1^2(\gamma_s + \kappa).
 \end{align}
Fig.~\ref{4} plots the comparison of the emission spectrums of cavity photon, obtained via Eq.~(\ref{eq7}) (blue circles) and by solving numerically the master equation (\ref{eq3}) (black solid curve) in the steady-state regime, versus the detuning $\Delta$. We see that the analytical result is full agreement with the numerical calculation. At resonant point, i.e., $\Delta=0$, the optimal average photon number and the equal-time second-order photon correlation function can be obtained
\begin{align}\label{eq12}
n_a^{(\Delta=0)}=&\frac{16\Omega^2\gamma_s\mu_1\mu_2 (8\mu_1\Omega^2\kappa + 2\gamma_s^3 + 5\gamma_s^2\kappa + 4\gamma_s\kappa^2 + \kappa^3)}{\kappa (\gamma_s + \kappa) (8\mu_1\Omega^2 + \gamma_s^2) (16\mu_1\Omega^2 + 2\gamma_s^2 + 3\gamma_s\kappa + \kappa^2)},\\
g^{(2)}_{(\Delta=0)}=&\frac{C(8\Omega^2\gamma_s + 8\kappa\Omega^2 + \gamma_s^3 + \kappa\gamma_s^2)(16\Omega^2 + 2\gamma_s^2 + 3\gamma_s\kappa + \kappa^2)}{D(8\Omega^2 + \gamma_s^2 + 3\gamma_s\kappa + 2\kappa^2)(\gamma_s^2 + 5\gamma_s\kappa + 6\kappa^2)},
\end{align}
where $C=C_1+192\Omega^4\kappa^2(\gamma_s + 2\kappa)C_2$ and $D=D_1(16\Omega^2 + 2\gamma_s^2+9\gamma_s\kappa+9\kappa^2)$,
\begin{align}\label{eq14}
C_1=&(8\Omega^2\kappa\gamma_s + 24\Omega^2\kappa^2)(4\gamma_s^3 + 18\gamma_s^2\kappa + 29\gamma_s\kappa^2 + 17\kappa^3), \\
C_2=&(4\gamma_s^3 + 12\gamma_s^2\kappa + 11\gamma_s\kappa^2 + 3\kappa^3) (\gamma_s^2 + 5\gamma_s\kappa + 6\kappa^2)^2,\\
D_1=&(8\Omega^2\kappa + 2\gamma_s^3 + 5\gamma_s^2\kappa + 4\gamma_s\kappa^2 + \kappa^3)^2.
\end{align}
\begin{figure}
  \centering
  \includegraphics[width=8cm]{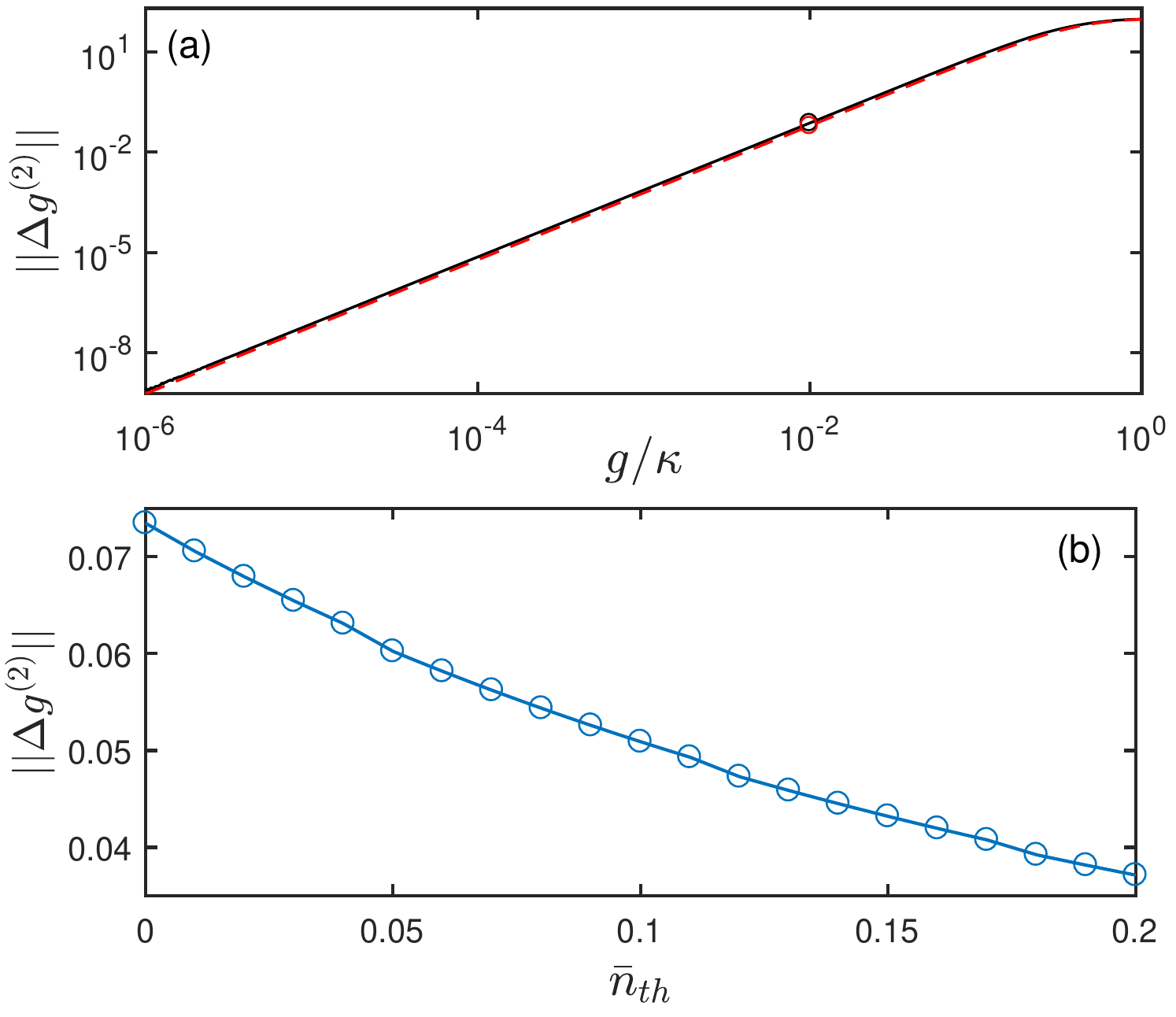}\\
  \caption{ (a) The mode $||\Delta g^{(2)}||$ versus $g$ for $\bar n_{th}=0$ (black solid curve) and $\bar n_{th}=0.05$ (red dashed curve), two circles correspond to $g=0.01\kappa$. (b) The mode $||\Delta g^{(2)}||$ versus the thermal average boson number $\bar{n}_{th}$ for $g=0.01\kappa$. The other system parameters used here are the same as in Fig.~\ref{5}(h)}\label{6}
\end{figure}

Furthermore, by calculating the master equation~(\ref{eq3}), we study the response of photon relations of the target system to the changes of interaction strength $g$, as shown in Fig.~\ref{5}. The first, second and third columns plot the response of cavity to the input of quantum lights from the central peak, the emission halfway between the central peak and right peak, and the right peak, respectively. $||\Delta n_a||=||n_a(g\neq0)-n_a(g=0)||$ and $||\Delta g^{(2)}||=||g^{(2)}(g\neq0)-g^{(2)}(g=0)||$ are, respectively, the differences of photon population and statistics in the JC model. Comparing the second row and the third row, photon statistics have higher sensitivity than photon population for different values of $g$. For later, we see that the value of $||\Delta n_a||$ is extremely low when the coupling strength is downed to a very small value in Fig.~\ref{5}(e). This is because there are very small photons scattered from source, as shown in Figs.~\ref{3} and ~\ref{4}. From the last row, it is shown that the value of $||\Delta g^{(2)}||$ in Fig.~\ref{5}(g) is greater than the other ones in Figs.~\ref{5}(h) and~\ref{5}(i) for a large value of $g$. However, we also find that, with the decreasing of interaction strength, $||\Delta g^{(2)}||$ in (h) has higher precision than the other ones. The reason is that quantum lights from the emission halfway between the central peak and each side peak are strongly correlated, and they have extremely strong quantum statistics characteristic~\cite{ref40}. We thus detect the weak interaction in the JC model by the quantum light from this frequency window. $||\Delta g^{(2)}||\approx0.0735$ is get for $g/\kappa=0.01$. For a weaker coupling strength, the difference is still obvious and could be used for experimental measurement.

In the discussion above about the dynamics of the coupling system, we assumed the temperature of the environment to be zero, i.e., the thermal average boson number $\bar{n}_{th}=0$. Now, we consider the system is in a non-zero temperature environment, the master equation can be replaced by
\begin{align}\label{eq17}
\frac {d\rho}{dt}=&i[\rho,H_s'+H_{t_1}']+\gamma_s\bar{n}_{th}\mathcal{L}[\sigma_s^\dag]+
\kappa\bar{n}_{th}\mathcal{L}[a^\dag]+\gamma\bar{n}_{th}\mathcal{L}[\sigma^\dag]\nonumber\\
&+\gamma_s(\bar{n}_{th}+1)\mathcal{L}[\sigma_s]+
\kappa(\bar{n}_{th}+1)\mathcal{L}[a]+\gamma(\bar{n}_{th}+1)\mathcal{L}[\sigma]\nonumber\\
&-(\bar{n}_{th}+1)\sqrt{\mu_2 \gamma_s \kappa} \{[a^\dag,\sigma_s \rho]+[\rho \sigma_s^\dag,a]\}\nonumber\\
&-\bar{n}_{th}\sqrt{\mu_2 \gamma_s \kappa} \{[a,\sigma_s^\dag \rho]+[\rho \sigma_s,a^\dag]\}.
\end{align}
We display the function of $||\Delta g^{(2)}||$ versus $g$ for $\bar n_{th}=0$ (black solid curve) and $\bar n_{th}=0.05$ (red dashed curve) in Fig.~\ref{6}(a). Here we excite the target system by quantum light from the emission halfway between the central peak and right peak. It is seen that our result is robust to the temperature. Here, the thermal occupancy $\bar n_{th}=0.05$ corresponds to the temperature $T=131.2$~mK, with $\bar n_{th}=[\texttt{exp}(\hbar\omega_c/K_B T)-1]^{-1}$. Moreover, Fig.~\ref{6}(b) presents the dependence of $||\Delta g^{(2)}||$ on the thermal average boson number $\bar{n}_{th}$ for $g=0.01\kappa$. We find that with the increase of the thermal average boson number, the values of $||\Delta g^{(2)}||$ show the trend of decrease. But the value can still reach to the order of $1\%$ even when $\bar{n}_{th}=0.2$.
\section{Detection of weak interaction in OMS by quantum light}

\begin{figure}
  \centering
  \includegraphics[width=7cm]{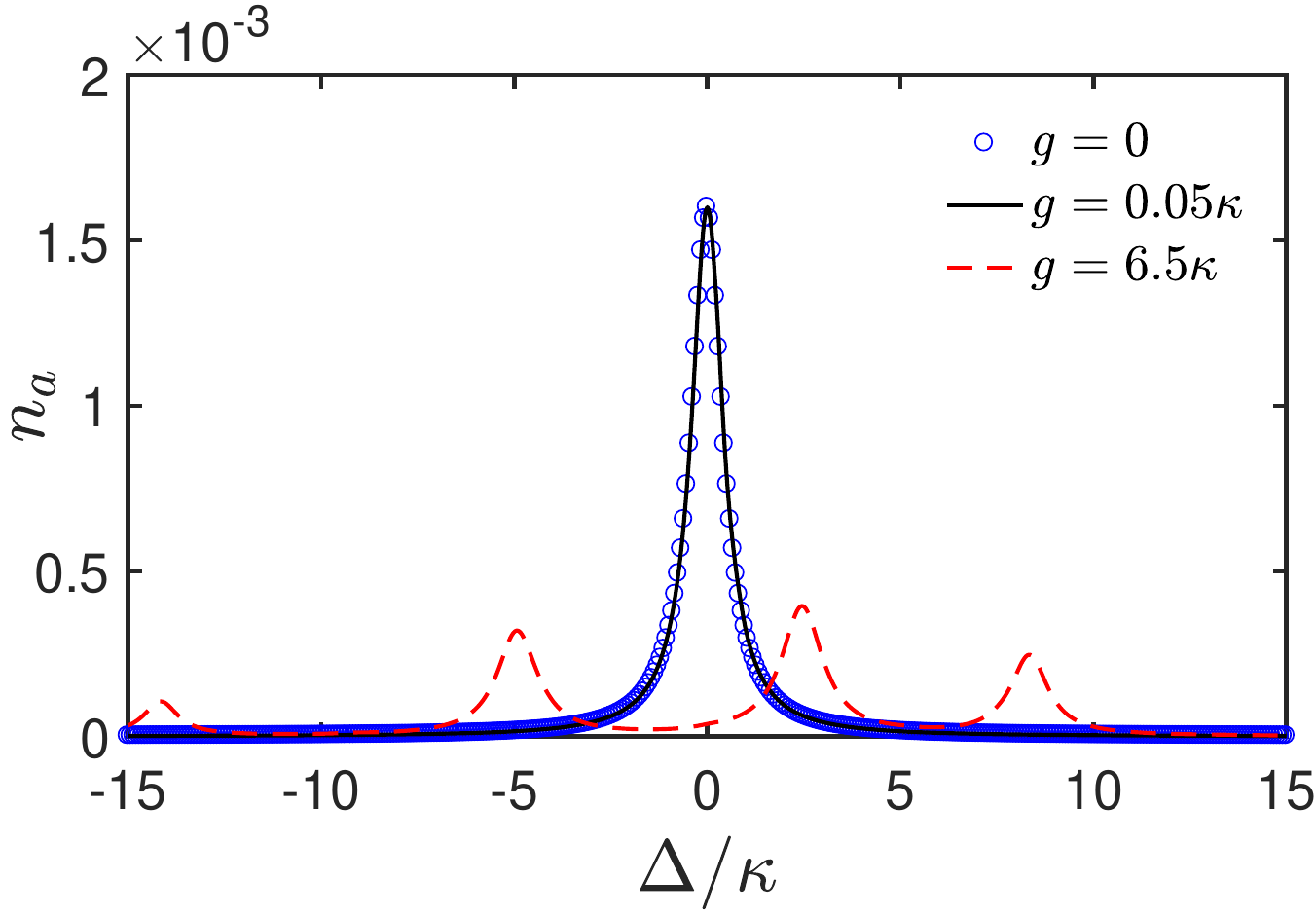}\\
  \caption{ Cavity mean photon number $n_a$ versus cavity-field detuning $\Delta$ when the OMS is driven by classical light field. The blue circles, black solid line and red dashed line in panels correspond to the coupling strength $g_m=0$, $g_m=0.05\kappa$ and $g_m=6.5\kappa$, respectively. The system parameters used here are: $\omega_m=5\kappa$, $\gamma_m=0.001\kappa$ and $\Omega=0.02\kappa$.}\label{7}
\end{figure}

\begin{figure*}
  \centering
  \includegraphics[width=15cm]{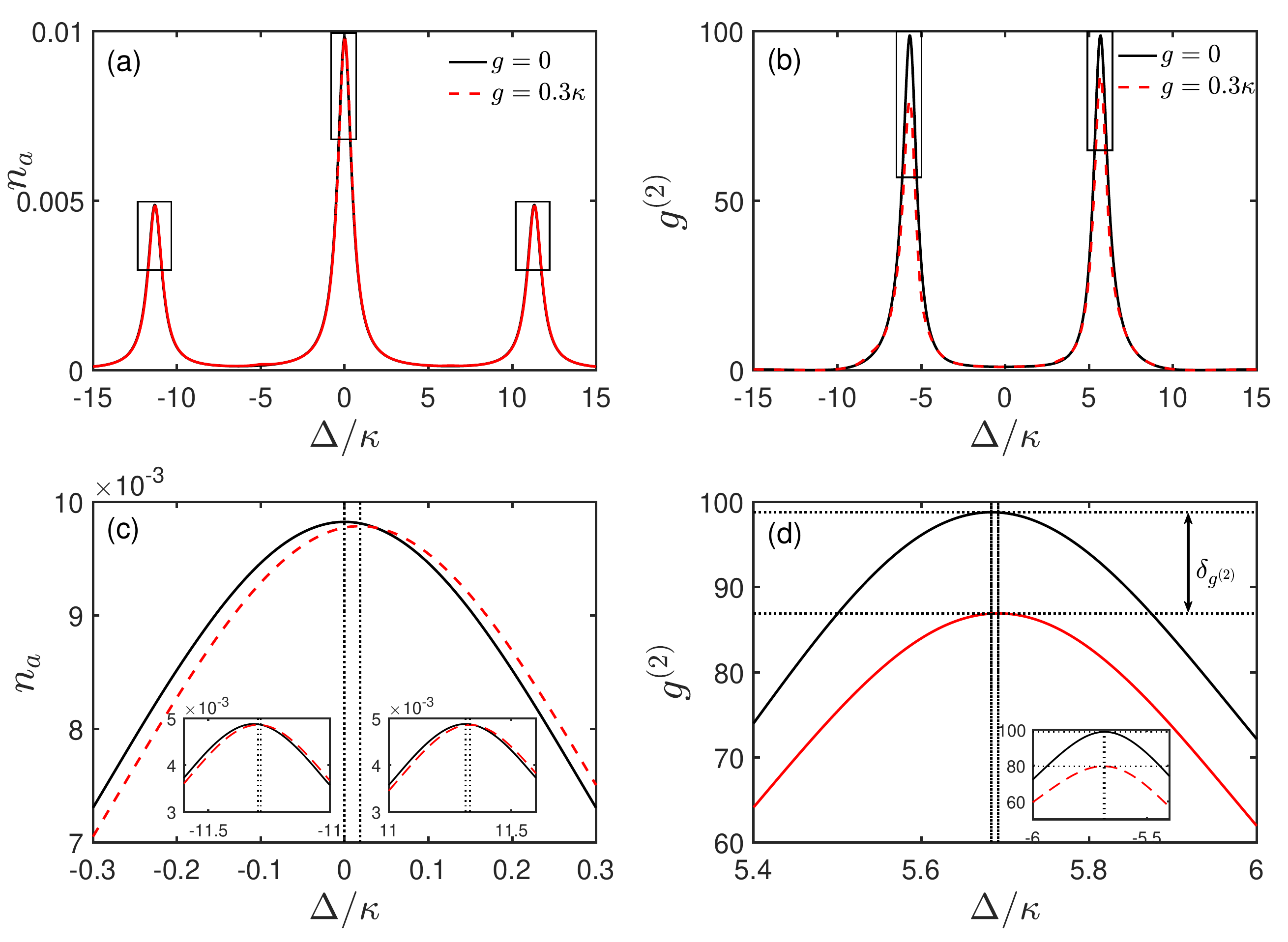}\\
  \caption{ Plot the photon emission spectrum (a) and photon statistics spectrum (b) when the Mollow triplet is scanned into the OMS. The black solid lines and red dashed lines in panels correspond to the coupling strength $g_m=0$ and $g_m=0.3\kappa$, respectively. (c) Enlarged view of the spectral region of the middle peak delimited by square in (a). Insets: Enlarged view the regions of the left and right emission peaks in (a). (d) Enlarged view of the regions delimited by squares in (b), and the left peak is enlarged in the inset. $\delta_{g^{(2)}}$ denotes the difference of the peak with $g_m=0.3\kappa$ and $g_m=0$. The system parameters used here are: $\omega_m=5\kappa$, $\gamma_s=0.02\kappa$, $\gamma_m=0.001\kappa$, $\Omega=8\kappa$, $\mu_1=0.5$ and $\mu_2=0.5$.}\label{8}
\end{figure*}
 For the OMS that coupling of a mechanical resonator to a cavity excited weakly by a coherent laser field. The shift of the cavity excitation spectrum can be seen when the system is in the strong coupling regime, as displayed in Fig.~\ref{7}~\cite{ref31}. Particularly, the peak has an obvious shift when the coupling strength is increased to a very large value. It is shown that the coupling strength $g_m$ in the strong coupling regime could be obtained by measuring the shift of the peak $\delta$~\cite{ref31,ref41,ref42}. However, the method is failed to detect the weak interaction in OMS. The reason is that the shift could be recovered by the width of spectrum line due to the high system dissipation. Therefore, it is hard to detect the weak interaction in this system by the classical light field. Just like the previous section, we further study the weak interaction between the cavity and mechanical modes by quantum light.

 We consider an OMS depicted in Fig.~\ref{1}(b), which consists of a mechanical resonator weakly coupled to a single-mode cavity driven by a quantum source. The source system is the same as the above section with Hamiltonian $H_s$. The OMS is set as a target system with the Hamiltonian
\begin{equation}\label{eq18}
H_{t_2}=\omega_c a^\dag a+\omega_m b^\dag b+g_m a^\dag a(b^\dag + b).
\end{equation}
Here, $b$ $(b^\dag)$ represents the annihilation (creation) operator of the mechanical mode with frequency $\omega_m$. $g_m$ is coupling strength between cavity and mechanical modes. We consider that the source system and OMS could be coupled in dissipative environments. In order to work out the dynamics of the coupled system, we investigate it in the framework of cascaded quantum system. Here the output field of the source system is set as the input field of the OMS via equations of motion. Assuming there is only one input channel $a^{in}(t)$ for the target. The cascaded system is then transformed into a frame rotating with $\omega_L$ and the full master equation is given by
\begin{align}\label{eq19}
\frac {d\rho}{dt}=&i[\rho,H_s'+H_{t_2}']+\gamma_s\mathcal{L}[\sigma_s]+\kappa\mathcal{L}[a]
+\gamma_m\mathcal{L}[b]\nonumber\\
&-\sqrt{\mu_2 \gamma_s \kappa} \{[a^\dag,\sigma_s \rho]+[\rho \sigma_s^\dag,a]\}.
\end{align}
with
\begin{align}\label{eq20}
H_{t_2}'=\Delta a^\dag a+\omega_m b^\dag b+g_m a^\dag a(b^\dag + b).
\end{align}
Here, $\gamma$ and $\gamma_m$ are, respectively, decay rates of the cavity and mechanical resonator in the target system, $\sqrt{\mu_2 \gamma_s \kappa}$ represents the coupling strength between source and target.
\begin{figure*}
  \centering
  \includegraphics[width=15.5cm]{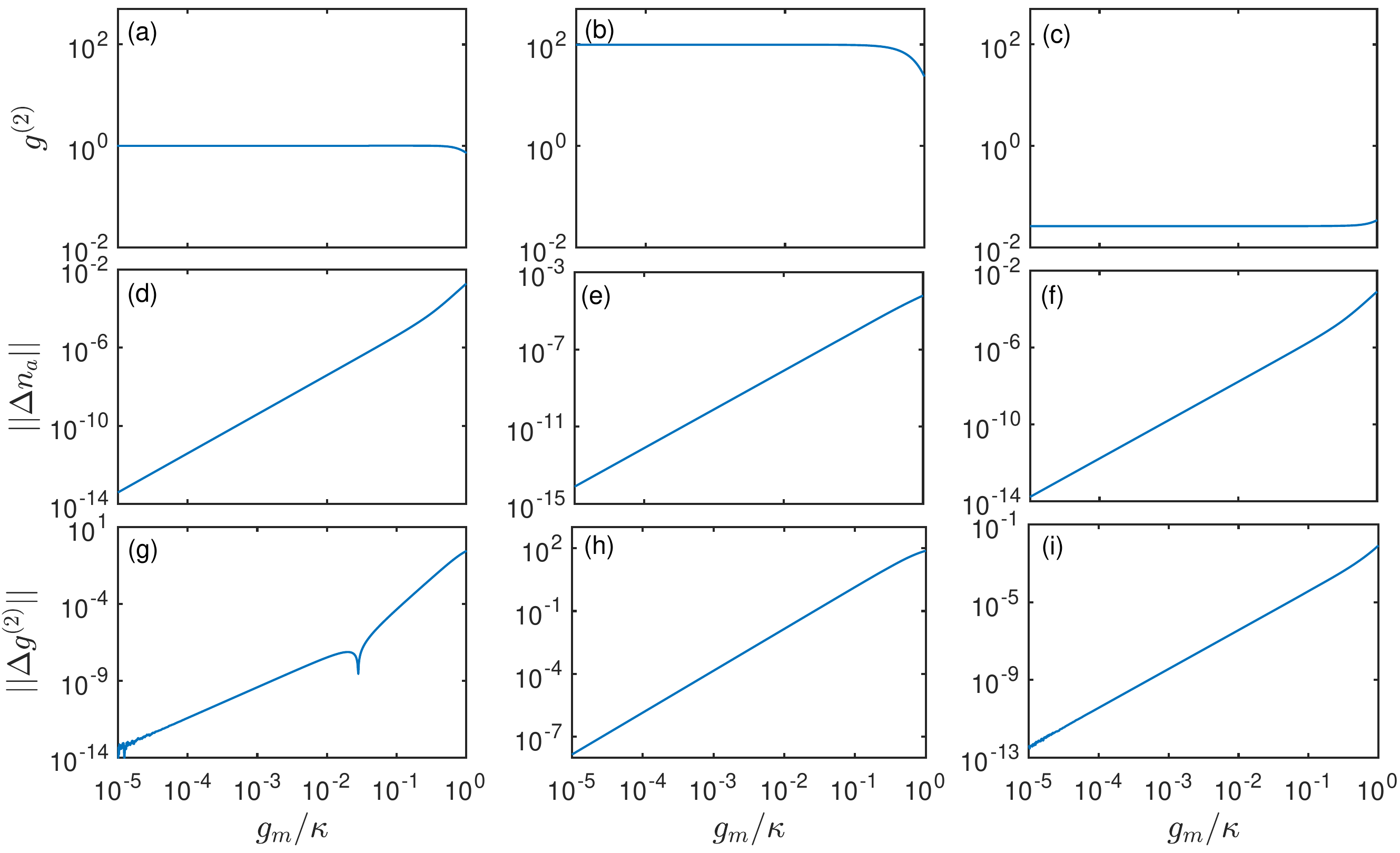}\\
  \caption{ The equal-time second-order correlation function $g^{(2)}$ (the first row), the norms $||\Delta n_a||$ (the second row) and $||\Delta g^{(2)}||$ (the third row) of transmitted light of the target cavity versus $g_m/\kappa$ for different input types of quantum lights. The first, second and third columns correspond to quantum lights from the central peak, the emission halfway between the central peak and right peak, and the right peak, respectively.  The system parameters used here are: $\omega_m=5\kappa$, $\gamma_s=0.02\kappa$, $\gamma_m=0.001\kappa$, $\Omega=8\kappa$, $\mu_1=0.5$ and $\mu_2=0.5$.}\label{9}
\end{figure*}

We now consider exciting the target cavity by the same quantum light as the previous section. The new emission and photon statistics spectroscopy are obtained when quantum lights from the source are scanned into the OMS, as displayed in Figs.~\ref{8}(a) and~\ref{8}(b), respectively. The new emission spectrum is mainly composed of three peaks, which also correspond to three forms of photon correlations, i.e., antibunched, bunched and superbunched. Here, both the population and the statistics of the source system are transferred to the target cavity, but with some deviations due to the presence of the interaction in OMS. Moreover, these spectrums are no longer symmetrical about $\Delta=0$. In Fig.~\ref{8}, the black solid curve and red dashed curve correspond to $g_m=0$ and $g_m=0.3\kappa$, respectively. From Fig.~\ref{8}(a), we see that two curves are basically coincident. Actually, there are small deviations in these peaks on an enlarged view as in Fig.~\ref{8}(c), showing that the entire emission spectrum has shifted. The shift arises from the coupling between cavity and mechanical modes. In Fig.~\ref{8}(b), we find that the black and red curves are obviously inconsistent. Fig.~\ref{8}(d) displays the enlarged view of the spectral regions delimited by squares in Fig.~\ref{8}(b). $\delta_{g^{(2)}}$ denotes the difference of the right peak with $g_m=0.3\kappa$ and $g_m=0$. It is seen that the large deviation of statistics is obtained in the case, and its value is much larger than that of photon population in the same parameter regime. We can also find that the statistical deviation of the left peak is not equal to that of the right peak, caused by the coupling between the cavity and mechanical modes in the OMS. To clarify the dynamics of the cascaded system, we obtain the exact solutions of photon population and photon statistics in the limiting case of $g_m=0$, whose results are the same as that of the above section ones. So we don't show them here.

 We have already mentioned above section that quantum light from Mollow triplet can be scanned over the JC model to probe the cavity-atom interaction in the weak coupling regime. We now consider using the same quantum light to detect the weak interaction between the cavity mode and mechanical resonator in OMS. To clarify the effect of the change of interaction strength $g_m$ on the coupling system dynamics, we have to solve the Eq.~(\ref{eq19}). In Fig.~\ref{9}, the first, second and third columns plot the responses of cavity to the input of quantum lights from the central peak, the emission halfway between the central peak and right peak, and the right peak, respectively. $||\Delta n_a||=||n_a(g_m\neq0)-n_a(g_m=0)||$ and $||\Delta g^{(2)}||=||g^{(2)}(g_m\neq0)-g^{(2)}(g_m=0)||$ correspond with the deviations of photon population and statistics in the OMS. In contrast to the second row and the third row, we see that the value of $||\Delta g^{(2)}||$ has higher sensitivity than that of  $||\Delta n_a||$ for different interaction strength $g_m$. For later, the value of $||\Delta n_a||$ is extremely low for a very small coupling strength. Because there are very small photons scattered from source, as displayed in Fig.~\ref{8}. Thus we can use the photon statistics spectroscopy to detect weak interaction in the OMS. From the last row, we see that, as the coupling strength increases, the deviations of $g^{(2)}$ in Fig.~\ref{9}(h) are always greater than that ones in Figs.~\ref{9}(g) and~\ref{9}(i). The reason is that quantum lights from the emission halfway between the central peak and each side peak are strongly correlated~\cite{ref40}, which lead to the emergence of high values of peaks in the statistics spectrum. Therefore, the large values of $||\Delta g^{(2)}||$ are obtained when exciting the target by quantum light from this frequency window, rather than that from other frequency windows. In Fig.~\ref{9}(h), we find that at $g_m/\kappa=0.01$ the difference $||\Delta g^{(2)}||\approx0.0136$. For a weaker value of $g_m$, a tiny value is obtained but it can still be used for experimental measurements.
\section{Discussion}
We envision an experiment for implementing our proposal in the near future. Firstly, regarding the source system, we consider a system that consists of an individual self-assembled (In, Ga) As/GaAs quantum dots (QDs) embedded in a high-quality microcavity~\cite{ref41}. The system is maintained at low temperature (131.2~mK) in a continuous-flow cryostat, and a polarization-maintaining single-mode optical fiber is brought close to the system edge. Then, we prepare a coherent laser field, made of the strong coherent pump field and vacuum field in an unitary mixer, coupling into the waveguide mode of the cavity through the fiber. Some fraction of the QDs resonantly coupled to the single-mode cavity and photons can be scattered from cavity~\cite{ref41,ref42,ref43}. Here, the system could be used to produce a tunable quantum light source.

Secondly, based on these experimental articles~\cite{ref2-1,ref44,ref45,ref45-1}, we construct the target system in which a two-level atom is trapped into a Fabry-P\'{e}rot cavity. Here, the cavity consisting of two highly reflective mirrors, separated by a distance $L=17.9$~mm. Its angular frequency $\omega_c\approx m\cdot\pi c/L\approx52.615$~GHz, where $m=1$ denotes the single-mode number. We chose a Rydberg atom with principal quantum number $n=50$ as the target two-level atom, whose transition frequency between two states $\omega_a\approx R/\pi\hbar n^3\approx52.615$~GHz, where $R$ is Rydberg constant. The Rydberg atom has a decay rate $\gamma\approx10$~Hz, i.e., the lifetime $\tau\sim0.1s$. Moreover, we place the system into a continuous-flow helium cryostat, which provides pre-cooling down to $T\approx131.2$~mK, reducing the bath occupancy of the 52.615~GHz single-mode cavity to $\bar n_{th}\approx0.05$. At this temperature, the microwave cavity has a total energy decay rate of $\kappa\approx10$~KHz, and the quality factor of the optical resonator $Q_c\approx5\times10^6$. Furthermore, we also construct an OMS system as the target system, which is made of a Fabry-P\'{e}rot cavity with a fixed macroscopic mirror and a movable micromechanical mirror, and the length of the cavity $L=17.9$~mm~\cite{ref2-1,ref45}. Similarly, the system is placed into the cryostat to pre-cool. Owing the speed of sound being much less than the speed of light, the mechanical resonance occurs at $\omega_m\approx50$~KHz with a quality factor $Q_m\approx5\times10^3$.

Thirdly, the light scattered from source can be scanned onto the target cavity to drive it to couple the atom (or mechanical resonator), this process could occur in a cascaded quantum system. Here, the output light of source system drives the target cavity via a waveguide supporting only a right-propagating mode~\cite{ref46,ref47}. In order to detect the light-matter interaction in the target system, we select the superbunched quantum light to excite the target system. This type of quantum light can be obtained by adjusting the frequency of the pump field to $\omega_L\approx106.9$~KHz. However, the average photon number in this frequency window is very small. Thus, this method improves the measurement accuracy but also needs to increase the number of pumping of the source.

Finally, the emitted photons from the target system were directed into the Hanbury Brow-Twiss (HBT) setups to measure $g^{(2)}$. The HBT setups comprised a beam splitter, two photon detectors from Micro Photon Devices and a photon counting system~\cite{ref48,ref49}.
\section{Conclusion}
In this theoretical work, we have studied the responses of the normal cavity QED system described by JC model and optomechanical system in weak coupling regime to the input fields of quantum lights from the source system. The quantum light can be scanned onto the cavity QED system to form new emission spectrum and statistics spectrum. The reason for the formations of spectrums is that both the population and the statistics of source system are transferred to the target cavity. But some deviations can be seen due to the presence of interaction between the cavity and atom (or mechanical oscillator) in the system. We have applied the new emission and quantum statistics spectrums to detect the weak light-matter interactions in cavity QED system and OMS when the strong dissipations are included. We have find that the photon statistics have higher sensitivity than photon population for different values of interaction strength. Moreover, the weak interactions can be read with high precision when exciting the target cavity by quantum light from the emission halfway between the central peak and each side peak, rather than that from other frequency windows. We have observed that the weak interaction can also be precisely measured even when the interaction is downed to a very small value. This work applied a new spectrum technique to detect the light-matter interaction in the cavity QED system and OMS under the weak coupling regime, which should advance the development of weak measurement and has potential applications in quantum information science.

\begin{acknowledgments}
This work is supported by National Key Research and Development Program of China (No. 2016YFA0301203); National Science Foundation of China (NSFC) (11374116, 11574104, 11375067).
\end{acknowledgments}

\end{document}